# Is the K-matrix approach adequate for describing overlapping resonances?

**V. Henner**[1,2], **T. Belozerova**[1]

[1]Department of Theoretical Physics, Perm State University, 614990 Perm, Russia

[2]Department of Physics and Astronomy, University of Louisville, KY, 40292, US

**Abstract**

The K-matrix method is widely used unitary parametrization for several resonances with the same quantum numbers. But in fact resonances in this approach are separated and do not overlap[1].

In the $K$-matrix approach the $S$-matrix is

$$S = \frac{I + i\rho K}{I - i\rho K}, \qquad (1)$$

where $\rho_i$ are phase-space factors. If operator $K$ is Hermitian, matrix $S$ is unitary and symmetric. With $S_{ij} = \delta_{ij} + 2iF_{ij} = \delta_{ij} + 2i\sqrt{\rho_i} T_{ij} \sqrt{\rho_j}$, for transition amplitudes it follows

$$T = (I - i\rho K)^{-1} K. \qquad (2)$$

A single pole parametrization $K = m_1 \Gamma_1 / (m_1^2 - s)$ leads to the standard Breit-Wigner (BW) function, $F = \rho m_1 \Gamma_1 / (m_1^2 - s - i\rho m_1 \Gamma_1)$, thus $m_1$ and $\Gamma_1$ are mass and width of a resonance in this simple case.

For several poles and channels, a common parametrization is [1]:

$$K_{ij} = \sum_\alpha^N \frac{m_\alpha \Gamma_\alpha \gamma_{\alpha i} \gamma_{\alpha j}}{m_\alpha^2 - s} + a_{ij} + b_{ij} s + \dots \qquad (3)$$

Parameters $m_\alpha$ and $\Gamma_\alpha$ are referred to as the nominal (or bare) mass and width, $\gamma_{\alpha i}$ as the coupling constant of the state $\alpha$ to channel $i$, polynomial as background – these statements are based on comparison with the BW function for an isolated resonance.

The feature that directly follows from expression (2) and therefore remains in any modification of the $K$-matrix method (for instance, amplitude left cuts in the resonant area effectively contribute to polynomials in (3)) is that scattering amplitudes have zero, or very close to zero values between the pole locations, $m_\alpha$. For example, for $M = 2$

$$T = \frac{1}{1 - \rho_1 \rho_2 D - i(\rho_1 K_{11} + \rho_2 K_{22})} \begin{pmatrix} K_{11} - i\rho_2 D & K_{12} \\ K_{21} & K_{22} - i\rho_1 D \end{pmatrix}, \qquad (4)$$

---

[1] *This short paper uses some results of the detailed text* [3]

$D = K_{11}K_{22} - K_{12}^2$ and for $N = 2$ the zero of equation $T_{12}(E) = 0$ is located between the poles $m_{1,2}$ and given (in absence of non-resonant terms in (3)) by a linear equation, for $N = 3$ by a quadratic equation. The zeros in $T_{11}$ and $T_{22}$ are slightly shifted from that location. This phenomena can be checked with the interactive software provided in supplemental material to this paper. If relatively narrow resonances are far from each other and background is neglected, the amplitudes approximately presented by a sum of isolated BW functions really become zero between the peaks. But in a situation when the states overlap, this feature of the $K$-matrix method can be considered as a defect - the resonances in $K$-matrix scattering amplitudes are always isolated and *actually do not overlap*. Same feature is translated on parametrizations that exploit the $K$-matrix for production amplitudes [2], $A_p = \sum_i (I - i\rho K)^{-1}_{pi} P_i$; vector $P_i$ has the same poles as $K$, it also can contain a polynomial. Nonzero reference level in mass projection of a Dalitz plot can only conceal this feature.

It is more useful to present simple examples rather than complex physical problem. Let us first consider a situation when resonances *overlap and not well resolved*, at least in some channels. Such a situation is presented in Fig. 1 for two resonances around 1.3 and 1.6 GeV having two common decay channels. The "data" are generated by drawing smooth curves, then discretized and randomized.

For comparison we also use the *unitary* BW approach [3] in which the unitarity is retained in the form

$$S_{ij} = e^{i(\beta_i + \beta_j)} \left[ \delta_{ij} + 2i\sqrt{\rho_i} \left( \sum_{r=1}^{N} e^{i\varphi_{ij}^{(r)}} \frac{m_r \Gamma_r^0 |g_{ri}| \cdot |g_{rj}|}{m_r^2 - s - im_r \Gamma_r(s)} \right) \sqrt{\rho_j} \right], \quad \Gamma_r = \sum_i^M \Gamma_{ri}. \quad (5)$$

Vectors $g_{ri} = e^{i\varphi_{ri}} |g_{ri}|$ are related with partial widths, $\varphi_{ij}^r = \varphi_{ri} + \varphi_{rj}$ are the interference phases. The widths $\Gamma_{ri}(s)$ can be energy-dependent which allows to take into account finite width effects and thresholds influence; $\Gamma_r^0 \equiv \Gamma_r(m_r^2)$. The background matrix $B_{ij} = \delta_{ij} e^{i\beta_i}$ is unitary. Without $B_{ij}$, the number of independent parameters in (5) is $N(M+1)$ – the same as in the pole terms in (4). In the BW scheme the production channels do not need special treatment – corresponding $g_{rp}$, having different physical nature, just may have different order of magnitude comparing to other $g_{ri}$.

We fit the data with $K$-matrix (4) and the BW formulas (5), $\rho_i(s) = \sqrt{(s - s_i)/s}$ (here we want to avoid unnecessary details like the barrier factors). In $T_{ij}$, continuation $\rho_2(s) \Rightarrow \pm i|\rho_2(s)|$ below $E_2$ can be used. In Figs. 1,2 we take $E_1 = 0.5$, $E_2 = 1.22$ GeV.

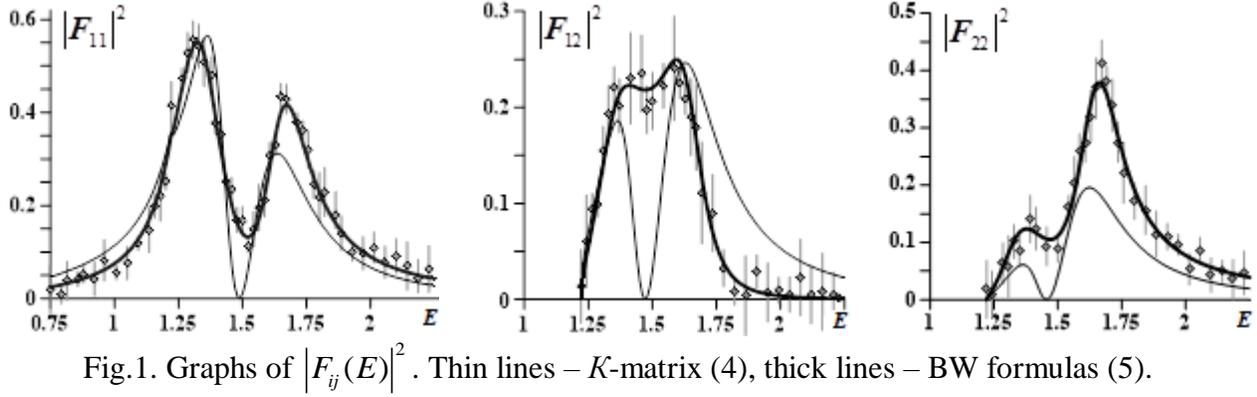

Fig.1. Graphs of $|F_{ij}(E)|^2$. Thin lines – K-matrix (4), thick lines – BW formulas (5).

The left side of Table 1 contains the K-matrix parameters, branching ratio is $B_{ri} = \gamma_{ri}^2$, $\sum_i \gamma_{\alpha i}^2 = 1$. The right side contains the BW parameters, $B_{ri} = |g_{ri}|^2 / \sum_{k=1}^{M} |g_{rk}|^2$, the partial width $\Gamma_{ri} = \Gamma_r B_{ri}$. By direct substitution of these $g_{ri}$ into (5) it can be checked that matrix $S(E)$ is unitary. The $\chi^2/d.o.f.$ in the BW and K fits are 0.53 and 4.75 respectively. The last value reflects the fact that the K-matrix inadequately describes the regions between the peaks.

**TABLE 1.** K-matrix and BW parameters (6 independent) for data in Fig. 1.

| K matrix parameters | | | | BW parameters | | | | | |
|---|---|---|---|---|---|---|---|---|---|
| $m_r$ | $\Gamma_r$ | $B_{r1}$ | $B_{r2}$ | $m_r$ | $\Gamma_r$ | $B_{r1}$ | $B_{r2}$ | $g_{r1}$ | $g_{r2}$ |
| 1.36 | 0.27 | 59 | 41 | 1.32 | 0.26 | 53 | 47 | –0.43–i0.03 | 0.41–i0.04 |
| 1.63 | 0.37 | 47 | 52 | 1.65 | 0.32 | 44 | 56 | 0.43-i0.03 | 0.49+i0.03 |

But when resonances are well resolved and do not overlap, both the K-matrix and BW descriptions lead to close results. This situation is presented in Fig. 2 in which the data have substantial dips between the resonances in all three channels. The quality of fits is practically the same in both methods, $\chi^2/d \approx 0.5$. The resonance parameters are collected in Table 2.

**TABLE 2.** K-matrix and BW parameters (6 independent) for data in Fig. 2.

| K matrix parameters | | | | BW parameters | | | | | |
|---|---|---|---|---|---|---|---|---|---|
| $m_r$ | $\Gamma_r$ | $B_{r1}$ | $B_{r2}$ | $m_r$ | $\Gamma_r$ | $B_{r1}$ | $B_{r2}$ | $g_{r1}$ | $g_{r2}$ |
| 1.30 | 0.28 | 45 | 55 | 1.30 | 0.19 | 49 | 51 | –0.39–i0.02 | –0.37–i0.03 |
| 1.60 | 0.12 | 27 | 73 | 1.60 | 0.09 | 18 | 82 | –0.11–i0.05 | –0.25+i0.06 |

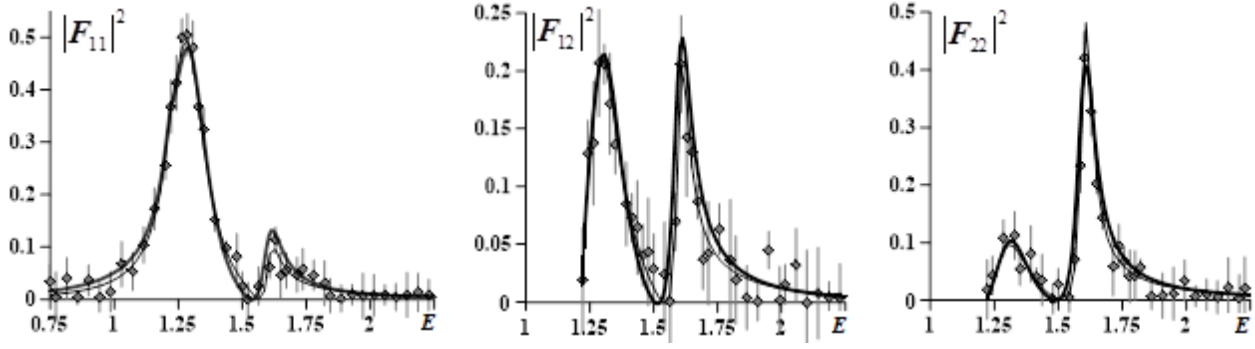

Fig.2. Graphs of $|F_{ij}(E)|^2$. Thin lines – *K*-matrix (4), thick lines – BW formulas (5).

In practice, resonances are accompanied by background. With polynomial in $K_{ij}$ the dips in amplitudes between the poles remain (this also can be checked with the accompanying software; an obvious restriction on polynomial coefficients is that resonances manifestation should retain). Also notice that polynomial terms in (3) do not allow to present background in the quantum mechanics form, $S_{ij}^B = e^{i(\beta_i + \beta_j)}$ in which even the number of parameters is different, for instance for two channels and energy-independent background two parameters $\beta_{1,2}$ versus three $a_{ij}$ in $K_{ij}$.

To conclude, it is important to be aware that actually overlapping resonances cannot be adequately described within the *K*-matrix parametrizations.

**Appendix:** *Examples of the K-matrix amplitudes*

Two states, two channels

For two poles in $K_{ij}$, the independent parameters are: six in the pole terms (their values can be changes by *using scroll bars*):

$$m_1, m_2, \Gamma_1, \Gamma_2, \gamma_{11}, \gamma_{21}$$

and six coefficients $A_{11}, A_{12}, A_{22}, B_{11}, B_{12}, B_{22}$ (their values can be *entered in the Table*).

*Enter the polynomial coefficients in the Table*

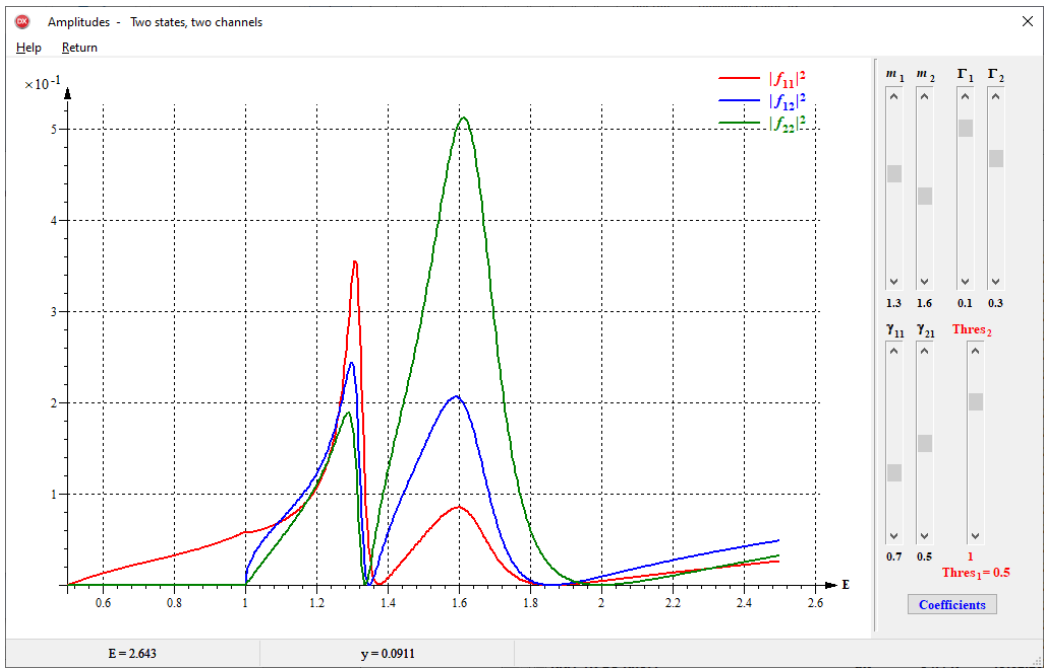

*To change the pole parameters and the 2<sup>nd</sup> threshold position use scroll bars*

Plots of $\left|F_{ij}\right|^2 = \rho_i \left|T_{ij}\right|^2 \rho_j$.

Three states, two channels

For three poles and two channels, the independent parameters are:
$m_1$, $m_2$, $m_3$, $\Gamma_1$, $\Gamma_2$, $\Gamma_3$, $\gamma_{11}$, $\gamma_{21}$, $\gamma_{31}$, , $A_{11}$, $A_{12}$, $A_{22}$, $B_{11}$, $B_{12}$, $B_{22}$.

*Enter the polynomial coefficients in the Table*

*To change the pole parameters and the 2<sup>nd</sup> threshold position use scroll bars*

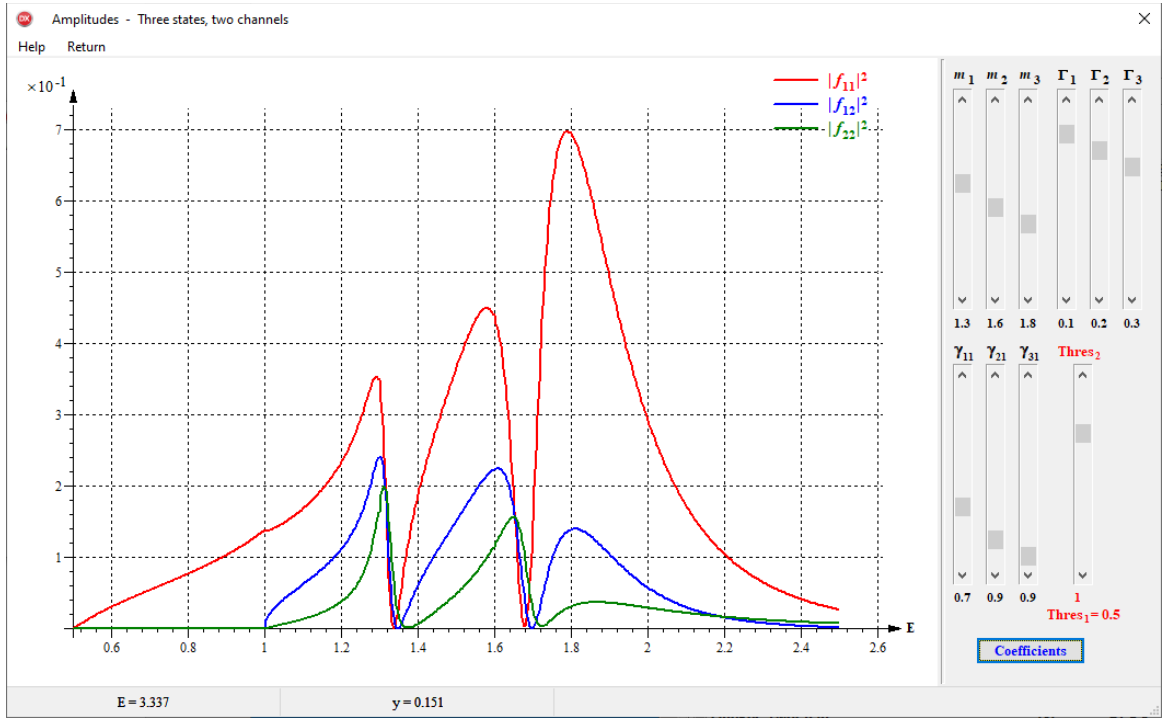

Three channels, two states

For three channels and two poles,

$$T = \frac{K}{1-iK} = \begin{pmatrix} K_{11} & K_{12} & K_{13} \\ K_{21} & K_{22} & K_{23} \\ K_{31} & K_{32} & K_{33} \end{pmatrix} \cdot \begin{pmatrix} 1-iK_{11} & -iK_{12} & -iK_{13} \\ -iK_{21} & 1-iK_{22} & -iK_{23} \\ -iK_{31} & -iK_{32} & 1-iK_{33} \end{pmatrix}^{-1}.$$

Eight independent parameters are (with normalization $\gamma_{a1}^2 + \gamma_{a2}^2 + \gamma_{a3}^2 = 1$):

$$m_1,\ m_2,\ \Gamma_1,\ \Gamma_2,\ \gamma_{11},\ \gamma_{12},\ \gamma_{21},\ \gamma_{22}.$$

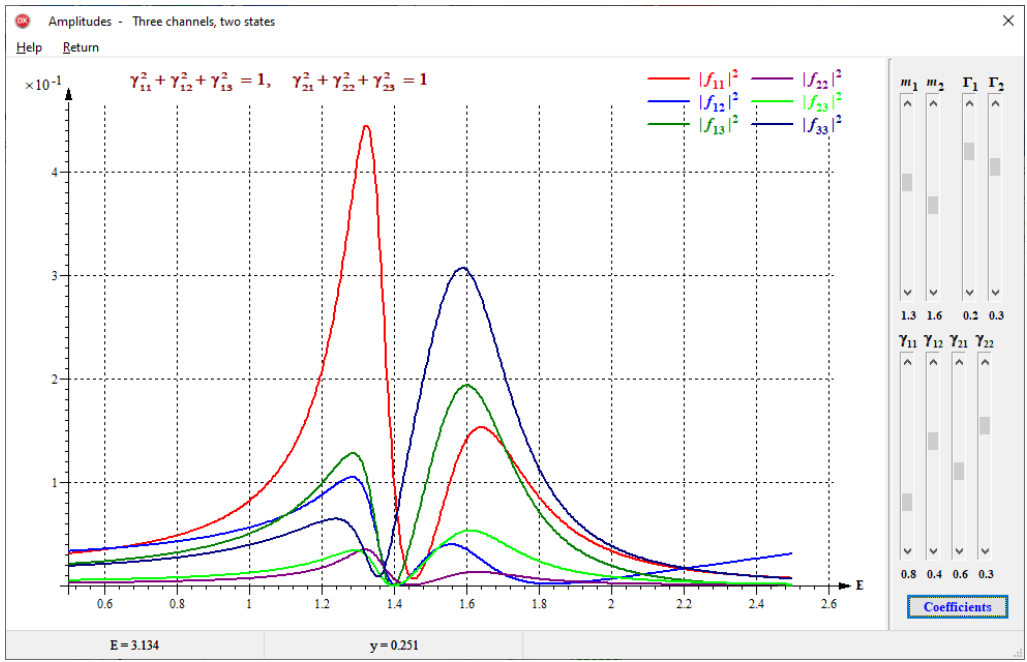